# Predicting Personality Traits from Physical Activity Intensity

Author

**Nan Gao**
RMIT University

**Wei Shao**
RMIT University

**Flora D. Salim**
RMIT University

**Guest Editors:**
Moustafa Youssef, E-JUST and Alexandria University

Fahim Kawsar, Nokia Bell Labs, UK and TU Delft, Netherlands

Abstract.

Call and messaging logs from mobile devices have been used to predict human personality traits successfully in recent years. However, the widely available accelerometer data is not yet utilized for this purpose. In this research, we explored some important features describing human physical activity intensity, used for the very first time to predict human personality traits through raw accelerometer data. Using a set of newly introduced metrics, we combined physical activity intensity features with traditional phone activity features for personality prediction. The experiment results show that the predicted personality scores are closer to the ground truth, with observable reduction of errors in predicting the Big-5 personality traits across male and female.

## INTRODUCTION

Traditional self-reported personality prediction had lots of limitations and relied too much on participants' objective answers, which can be time-consuming and less accurate. Past research has shown that it is possible to predict a human's personality through historical records of mobile data, such as call, message, app usage, and location logs [1-3].

Studies have indicated that physical activity intensity has a strong correlation with human personality [4]. Accelerometers have been widely applied in various devices such as mobile phones and fitness wristbands to detect human physical activity intensity [1]. To predict personality traits through mobile phones, researchers mainly focused on exploring phone activities or app usage.

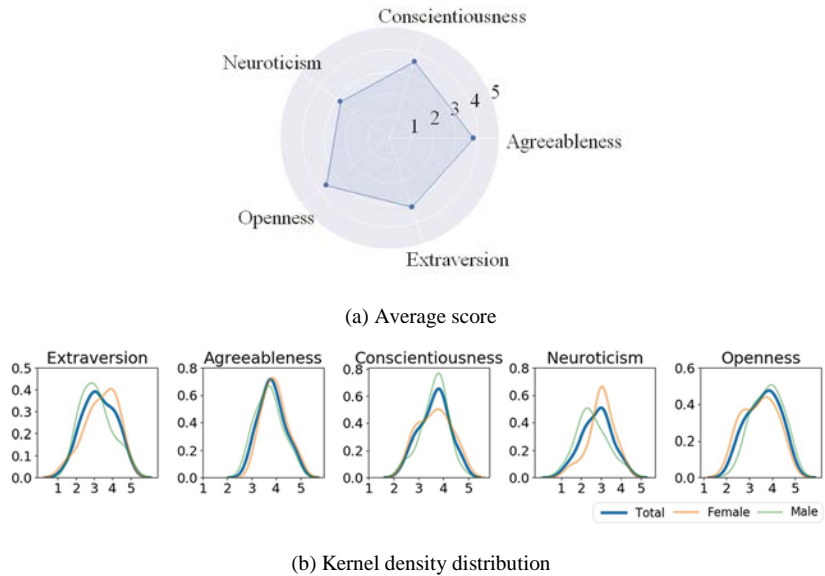

(a) Average score

(b) Kernel density distribution

Fig. 1. Big-5 personality traits in the dataset

However, nobody has taken advantage of combining phone activity data with physical activity intensity data from accelerometer sensors. In this paper, we assume that we can predict human personality by exploring the participants' phone activity and physical activity intensity. Besides, since female and male usually have different activity patterns, we run the experiments separately for different genders.

The Big-Five personality framework is one of the most important measurements of personality traits [5, 6], which consists of five dimensions: Extraversion, Agreeableness, Conscientiousness, Neuroticism, Openness to Experience (Openness). Extraversion reflects the degree of energetic, sociable and talkative. Openness represents the tendency to be curious and inventive. Agreeableness usually means the potential to be friendly and compassionate instead of being suspicious and hostile to others. Conscientiousness shows the tendency to be organized, efficient and careful. Neuroticism displays the tendency to be nervous and sensitive instead of being confident and secure. Fig. 1a shows the average score for five different personality traits in our data set (see Section 3.3 for more details).

In this paper, we are the first to combine physical activity intensity data and phone activity data to predict human personality. We proposed several important metrics based on *Diversity*, *Dispersion*, and *Regularity*. Then we categorized these features based on different temporal factors and genders. We applied Support Vector Regression (SVR) to build a prediction model of human personality traits. Experiment results showed that using features from human physical intensity through accelerometer data can improve the prediction accuracy. In addition, the prediction performance improvement is different between male and female when activity intensity is taken into account.

The contributions of the paper are as follows:

- We predict human personality traits for the first time by combining physical activity intensity data with traditional phone activity data.

- We proposed novel metrics based on different categories: *Diversity*, *Dispersion* and *Regularity*, and identified some significant associations between human physical activity intensity and self-reported personality traits.

- We found that the features describing physical activity intensity can improve the performance of personality prediction, with observable reduction of errors across the male and female group.

To the best of our knowledge, this is the first attempt to combine human physical activity intensity data with traditional phone activity data for personality traits prediction. Our results lay the foundation for the future study on this topic. The rest of the paper is organized as follows. Section 1 introduced the basic concepts of personality prediction, limitation of current research and our assumptions. Section 2 discussed the current related works on personality prediction. In section 3, we introduced the data set, proposed several activity behavior metrics based on different categories and reported the Big-5 personality ground truth. In section 4, we conducted association analysis based on extracted features and made a prediction through regression analysis. Section 5 gave conclusions and future work.

# RELATED WORK

Machine learning techniques have been applied successfully on sensor data such as for predicting human's mobility [7], identity [8], activities, transportation modes and complex behaviors [9]. Users' personality traits can be predicted through various media applications. Online social networking sites has been used to reflect user's personality such as Facebook profiles [10] and Facebook messages [11]. Nhi et al. proposed a personality mining framework to exploit the information from videos (e.g., Youtube clips), which includes visual, auditory and textural perspectives [12]. Xin et al. demonstrated the relationships between active users' micro-blogging behaviors and personality traits [13]. Some other works show that it is possible to estimate the personality traits of users through exploring the mobiles usage behaviors that are inferred from mobile data such as call logs, applications logs, Bluetooth logs and message logs [14].

Cabrera-Quiros et al. [15] recognized self-assessed personality during crowded mingle scenarios using accelerometers and proximity sensors embedded in wearable devices alone. Although they considered the physical activity of each person, their research requires people to wear the same wristband in the specified scenarios, which is not applicable in our daily life. Recently, Weichen et al. [3] predicted personality traits through mobile sensings such as ambient sound, ambient voice, physical activity, phone activity, etc. However, they only computed the *sedentary duration within every hour* to represent the pattern of physical activity, which is simple and naïve since they did not consider the whole physical activity intensity distribution. For phone activity, they merely used *the number of phone lock/unlock events* and *unlock duration* to estimate the phone usage, while did not consider the diversity/regularity/dispersion of phone contacts.

Mobile logs (phone and message activity) are easily accessible and has been used for efficient personality prediction [2]. To the best of our knowledge, there has been no application to infer personality by combining traditional mobile activity with physical activity intensity, which is proved to have a strong association with human personality [4]. Physical activity intensity can be estimated through data from accelerometers [16], which have been widely deployed in multiple devices such as mobile phones, fitness wristband, etc.

# METHODOLOGY

## Participants and Procedure

In the research, we exploited a dataset including 55 participants living in a young-family residential community adjacent to a major research university in North America between March 2010 and July 2011 [17]. Each participant was equipped with an Android OS based mobile phone running with a sensing software *Funf* [17] designed for periodically collecting mobile data. The software operates in a passive way and thus does not influence users' normal usage on the mobile phone. At the initial stage of data collection, each participant needs to complete a personality survey, and Big-5 scores can be calculated through [6]. After removing participants who do not have a complete response to the Big-5 survey, our final sample was composed of 52 participants (27 female, 25 male).

For this dataset, we will mainly focus on users' activity data which includes phone activity and physical activity data. Phone activity data such as call and text messages received/sent has been widely used in personality prediction [2]. Also, physical activity data inferred from accelerometers has been proved to have a strong association with peoples' personality [4]. Thus, in this research, we limit the scope of the study to the participants' call logs, text message logs, accelerometer logs, which is easily accessible for the future mobile data collection.

For accelerometer logs, raw 3-axis measurements are sampled at 5 Hz rate during 15 seconds in every 2 minutes. Participants were not constrained in the way they should carry the phone. For call logs and message logs, the human-readable texts are captured as hashed identifiers. For more details about the dataset, please see [17].

## Activity Behavior Metrics

Human's personality can be evaluated through Big-5 model, which consists of five major dimensions of personality traits: Openness (i.e., the tendency to be curious and inventive), Extraversion (i.e., the tendency to be energetic and energetic), Agreeableness (i.e., the tendency to be friendly and compassionate), Conscientiousness (i.e., the tendency to be organized and efficient) and Neuroticism (i.e., the tendency to be nervous and sensitive).

To better understand the patterns of human activity in their daily lives, we computed several metrics that could meaningfully tell the differences between personality traits. The metrics are divided into three categories: dispersion, diversity, and regularity. We used these metrics to evaluate the participants' phone activity and physical activity.

Phone activity includes call and messages interactions, which will be computed separately based on the metrics. For physical activity, we first partition the raw accelerometer data in one day into 24-hour periods and process it through an hourly fashion. Then we use the *Mean Amplitude Deviation* (MAD) [18, 19] across each hour to assess the intensity of physical activity.

$$MAD = \frac{1}{n}\sum |r_i - \bar{r}|, \quad (1)$$

where $n$ is the number of accelerometer data samples in each time period, $r_i$ means the resultant acceleration at the $i$th time stamp, and $\bar{r}$ represents the mean resultant value across the time period. $r_i$ can be calculated through

$$r_i = \sqrt{x_i^2 + y_i^2 + z_i^2}, \quad (2)$$

where $x_i, y_i, z_i$ represents the $x, y, z$-direction of the raw acceleration signal. Next, we compute the following metrics for assessing activity behaviors as follows.

**Dispersion of Activity Behaviors** depicts how sporadic activity behavior is. In our research, *Standard Deviation* (SD) is used to evaluate the dispersion of people's phone activity and physical activity intensity. Since people tend to have different activity patterns in the different time (i.e., more physical activity in weekends or fewer phone calls in the night), we compute the SD for three-time stages (daytime, evening and night) across weekdays and weekends during the data collection period.

**Diversity of Activity Behaviors** means the state of being diverse for users' activity. *Shannon Entropy* measures the amount of disorder in a system, which can be used to measure the diversity of users' contacts:

$$S = -\sum_{i=1}^{n} F_i \log F_i, \quad (3)$$

where $F_i$ means the frequency that user $s$ interacts with $i$ of all contacts $n$. Higher entropy represents user $s$ interacts equally with plenty of contacts and lower entropy happens when the user mostly interacts with specified contacts. *Shannon Entropy* is used to evaluate the diversity of phone activity in this study.

**Regularity of Activity Behaviors** means the state of being regular patterns. We propose the *Regularity Index (RI)* based on [3] to calculate the difference between specified time periods $T$ in two different days. Firstly, we rescale the data for each participant to [-1, 1], where -1 corresponds to the minimum value in the original data and 1 corresponding to the maximum value. The regularity index is positive if the values are close, and negative if they are not similar. Then, we define the $RI$ of the time period $t$ between day $i$ and day $j$ as:

$$\forall (i,j) \in S, RI_{i,j}^T = \frac{1}{T}\sum_{t=1}^{T} x_t^i x_t^j, \qquad (4)$$

TABLE 1. Overview of the Big-5 Scores for Total/Male/Female Participants

| Gender | Personality Traits | Mean | Std. Dev. | Median | Min | Max |
|---|---|---|---|---|---|---|
| Total | Extraversion | 3.26 | 0.86 | 3.13 | 1.50 | 4.88 |
| | Agreeableness | 3.83 | 0.52 | 3.78 | 2.78 | 5.00 |
| | Conscientiousness | 3.64 | 0.58 | 3.78 | 2.44 | 4.67 |
| | Neuroticism | 2.79 | 0.74 | 2.88 | 1.13 | 4.25 |
| | Openness | 3.61 | 0.70 | 3.70 | 2.20 | 4.90 |
| Female | Extraversion | 3.38 | 0.87 | 3.63 | 1.50 | 4.63 |
| | Agreeableness | 3.95 | 0.50 | 3.89 | 3.11 | 5.00 |
| | Conscientiousness | 3.65 | 0.64 | 3.67 | 2.67 | 4.67 |
| | Neuroticism | 3.00 | 0.65 | 3.00 | 1.38 | 4.13 |
| | Openness | 3.44 | 0.72 | 3.50 | 2.20 | 4.60 |
| Male | Extraversion | 3.13 | 0.84 | 3.00 | 2.00 | 4.88 |
| | Agreeableness | 3.71 | 0.54 | 3.67 | 2.78 | 4.78 |
| | Conscientiousness | 3.62 | 0.53 | 3.78 | 2.44 | 4.67 |
| | Neuroticism | 2.56 | 0.77 | 2.38 | 1.13 | 4.25 |
| | Openness | 3.80 | 0.64 | 3.90 | 2.50 | 4.90 |

Where $S$ is the set to two time period pairs, $x_t^i$ and $x_t^j$ means the rescaled value at hour $t$ in the time period $T$. We compute the average $RI$ values from every possible pair within the following sets: (1) within all days, (2) within weekdays, (3) within weekends, (4) within weekday daytime, (5) within weekday nights, (6) within weekday evenings, (7) within weekend daytime, (8) within weekend evenings, (9) within weekend nights. $RI$ is used to evaluate the regularity of phone activity and physical activity in this study.

To prove the advantages of extracted physical activity features and make the comparison fair, we also obtained some traditional phone activity features based on previous literature [2], including 'average of inter-event time', 'variance of inter-event time', 'response rate', 'response latency', 'percent during the night' and 'percent initiated'. Table 2 summarizes the features used in our study.

## Big-5 Personality Ground Truth

We used the self-reported Big-5 results from the participants as the ground truth for different personality traits. The scores were computed from 52 questions related to different personality traits [20], and the score is from 1 to 5 where 1 means the lowest score and 5 indicates the highest score of the personality trait. Fig. 1b shows the distribution of five personality traits based on gender differences.

The descriptive statistics results (mean value, standard deviation, median value, minimum value, and maximum value) for the entire population and different gender groups are given in Table 1. For the entire population, the average score of different personality traits is close to 3. It can be observed that the average score of agreeableness is around 4, followed by conscientiousness, openness, extraversion, and neuroticism. Agreeableness trait has the lowest standard deviation, which means that the agreeableness scores of most participants are very close.

Interestingly, we found that female and male had different distribution patterns in five personality traits. Especially, female usually have higher neuroticism score than male (t-test p-value=0.03). This leads us to believe that the female in our population sample are more sensitive and emotional than male. Furthermore, the male seems to have higher openness scores than female, which indicates that most males are likely to be curious while the female tends to be cautious.

# ASSESSING PERSONALITY USING ACTIVITY PATTERNS

## Feature Analysis

We extracted features based on the introduced metrics and different time spans in section 3.2 (see Table 2). We define the daytime period starts from 9:00 am to 6:00 pm, the evening period starts from 6:00 pm to 12:00 am and the night period starts from 12:00 am to 9:00 am.

TABLE 2. Description of the Extracted Features

| Feature | Features computed | Data |
|---|---|---|
| *Dispersion* | STD on the number of interactions for all days | call, message, c&m |
| | STD on physical activity intensity for all days on daytime/evening/night | accelerometer data |
| | STD on physical activity intensity for weekdays on daytime/evening/night | accelerometer data |
| | STD on physical activity intensity for weekends on daytime/evening/night | accelerometer data |
| | STD on physical activity magnitude for all days | accelerometer data |
| *Diversity* | Entropy of total contacts for all days | call, message, c&m |
| | Entropy of total contacts for weekdays | call, message, c&m |
| | Entropy of contacts in sent box for all days | call, message, c&m |
| | Entropy of contacts in sent box for weekdays | call, message, c&m |
| *Regularity* | Average RI of number of interactions for all days | call, message, c&m |
| | Average RI of physical activity intensity | accelerometer data |
| | Variance of RI for the number of interactions on the daytime/evening/night | call, message, c&m |
| | Variance of RI for physical activity intensity on the daytime/evening/night | accelerometer data |
| *Basic* | Total number of interactions for all days/weekdays | call, message, c&m |
| | Average physical activity intensity for all days on the daytime/evening/night | accelerometer data |
| | Average physical activity intensity for weekdays on the daytime/evening/night | accelerometer data |
| | Average physical activity intensity for weekends on the daytime/evening/night | accelerometer data |
| | Average inter-event time for all days | call, message, c&m |
| | STD on inter-event time for all days | call, message, c&m |
| | Contacts to interactions ratio for all days | call, message, c&m |
| | Response rate for all days | call, message |
| | Response latency for all days | call, message |
| | Percent during the night for all days | call |
| | Percent initiated for all days | call, message, c&m |

Since most features except for entropy metrics were strongly positively skewed, we applied log transformation for these features before conducting correlation analysis. Then, the *Pearson Correlation Coefficient* (PCC) is calculated between extracted activity features and Big-5 personality, which is widely applied to measure the correlation between variables in the psychology field. The value of PCC is between -1 to 1, where 1 represents the total positive linear correlation, 0 means no linear correlation and -1 indicates the total negative linear correlation. Table 3 show the top 3 useful features to predict Big-5 personality scores for total participants and female/male participants, where (+) represents the positive correlation and (−) means the negative correlation with the personality traits. In Table 3, we also list the PCC value for each useful feature. Next, we will discuss them in detail.

TABLE 3. Most useful Features to Predict Personality Traits (Female and Male Population)

| Personality | Gender | Top-3 Features |
|---|---|---|
| Extraversion | Female | (+0.55) Average physical activity intensity on weekend evening<br>(-0.46) Average inter-event time of messages<br>(-0.40) Response latency of messages |
| | Male | (0.44) Entropy of call & messages<br>(+0.25) Average physical activity intensity on weekday evenings<br>(-0.25) RI of physical activity intensity on weekday evenings |
| | Total | (-0.30) RI of physical activity intensity on weekday evenings<br>(+0.26) Entropy of contacts of call and messages<br>(-0.23) STD of physical activity intensity on weekday daytime |
| Agreeableness | Female | (+0.35) Number of outgoing calls<br>(-0.37) Percent of initiated calls<br>(-0.31) RI of physical activity intensity on weekday nights |
| | Male | (+0.47) Average physical activity intensity on weekday evenings<br>(-0.43) RI of physical activity intensity on weekday evening<br>(+0.39) Percent of initiated messages |
| | Total | (-0.33) RI of physical activity intensity on weekday evenings<br>(+0.26) Average physical activity intensity on weekends<br>(+0.23) Average physical activity intensity on weekday evenings |
| Conscientious-ness | Female | (+0.42) RI of physical activity intensity on weekend daytime<br>(+0.35) Entropy of calls and messages<br>(+0.21) Average physical activity intensity on weekend evenings |
| | Male | (+0.51) Entropy of call and messages<br>(-0.35) RI of physical activity intensity on weekend evening<br>(+0.34) Number of messages |
| | Total | (+0.44) Entropy of call & messages<br>(+0.27) Total number of messages<br>(+0.20) Average physical activity intensity on weekend evenings |
| Neuroticism | Female | (+0.44) RI of physical activity intensity on weekend nights<br>(+0.42) Entropy of calls<br>(+0.36) RI of physical activity intensity on weekday nights |
| | Male | (-0.30) RI of physical activity intensity on weekday nights<br>(+0.21) Entropy of call & messages<br>(-0.20) RI of physical activity intensity on weekend nights |
| | Total | (+0.27) Entropy of calls<br>(+0.25) Response latency of messages<br>(-0.24) STD of physical activity intensity on weekend daytime |
| Openness | Female | (−0.27) Total number of calls<br>(−0.22) RI of physical activity intensity on weekday evenings<br>(−0.20) Average physical activity intensity on weekday nights |
| | Male | (-0.32) Total number of calls<br>(+0.29) STD of physical activity intensity on weekday evenings<br>(+0.19) Percent of initiated calls |
| | Total | (-0.32) Total number of calls<br>(+0.26) STD of physical activity intensity on weekday evening<br>(+0.21) Average Inter-event time of calls |

**Extraversion.** The Regularity index of physical activity intensity for weekday evenings is negatively associated with the extraversion trait. This suggests that people who have higher extraversion score, usually do not follow similar patterns on the weekday night. The high entropy of contacts means that they tend to interact with different people randomly, which is in accordance with our experience in daily life.

**Agreeableness.** Similar to the extraversion trait, the people with high agreeableness usually have low regularity index of physical activity for weekday evenings since they may be sociable. They also tend to have more activities on weekends and weekday evenings. It is highly likely that a friendly and compassionate female usually have more outgoing calls.

**Conscientiousness.** We find that both for female and male with high conscientiousness score, they tend to have high entropy of contacts. It tells us that people who are more organized and efficient tend to contact different people and don't usually connect to the same people. Also, organized people may have high activity intensity on weekend evenings because they have already planned it before and get everything well-prepared.

**Neuroticism.** We find that the regularity index of physical activity intensity on weekday and weekend nights for the female is positively correlated with neuroticism. This leads us to believe that the female who is sensitive seems to have regular physical activity in the night (after 12:00 am). Interestingly, these same features for the male group are negatively correlated with neuroticism, which displays the difference between men and women.

**Openness.** We find the total number of calls is negatively correlated with the openness trait. In addition, the average inter-event time of calls is positively correlated with the openness score. That is to say, individuals who have fewer phone calls and a longer period between each call tend to be more inventive and curious.

## Prediction Analysis

Personality prediction is commonly regarded as a regression problem and the value of the score ranges from 1 (lowest) to 5 (highest) of each personality trait. Although the personality score can be divided into several classes (e.g., high, medium, low) using a certain threshold, researchers have proven that it is not a good practice to determine people's psychology characteristics. Most classification models showed a pretty low prediction accuracy of around 49% to 63% [2]. Thus, in the study, we will use the regression model to predict personality traits.

Support Vector Regression (SVR) with Radial Basis Function (RBF) kernel is chosen to predict the Big-5 personality scores. SVR has been applied in various fields and can deal with high dimensional data and automatically model non-linear relationships. Since there exist noticeable dissimilarities for personality scores among different genders and the key features are not the same, we conduct the prediction by choosing the best regressors for the entire population, male and female group separately.

### Baseline and Evaluation

Through the literature review, we found that most researchers used the random chance or majority class selection method as the baseline for Big-5 personality prediction [1], [2], [3]. However, in our research, we aim to improve prediction performance by combining human physical activity features with traditional phone features. Thus, it does not make much sense to compare our model with the random chance or majority class selection as the personality traits are hard to be predicted from only one kind of data. In the experiment, personality prediction with only phone activity data (call logs and messages logs) with state-of-art metrics (introduced in Section 3.2) is considered as the baseline model in the experiment.

For evaluation, we adopted the leave-one-out validation method because it usually shows the best performance when estimating the model from a small data set. Using leave-one-out method, we calculated the average value for MAE and MSE for each personality traits.

We validated our model with the Mean Absolute Error (MAE) and Mean Squared Error (MSE).

$$MAE = \frac{1}{n}\sum_{i=1}^{n}|y_{true} - y_{pred}|, \quad (5)$$

$$MSE = \frac{1}{n}\sum_{i=1}^{n}(y_{true} - y_{pred})^2, \quad (6)$$

where $n$ represents the number of samples, $y_{true}$ means the true personality scores and $y_{pred}$ means the predictive value of personality scores. The MAE and MSE can describe the goodness of predictions compared with the ground truth of personality score. The closer the MAE and MSE are to 0, the more successful the modal forecast.

## Discussion

Table 4 displays the performance of our prediction model based on the extracted features from call logs, message logs, and raw accelerometer logs. With the observable reduction of errors, our model has a better performance than the baseline model for all personality traits. The predicted Big-5 scores are highly correlated with the ground truth.

Upon comparison of the MAE and MSE between our model and the baseline model, it is interesting to note that the conscientiousness, neuroticism, extraversion were the personality traits that were best predicted in our predictive model. For the entire population, the model predicting conscientiousness score achieves 0.249 of MAE, which is 0.148 (37.28%) lower than the baseline model. For the female group, the model predicting neuroticism score achieves 0.425 of MSE, which is 0.129 (23.29%) lower than the baseline model. In the meantime, the MSE of extraversion score for the female group is 0.128 (17.56%) lower than the baseline.

Table 4. Prediction Performance for Total/Male/Female Participants

| Group | Big-5 Traits | MAE | | MSE | |
|---|---|---|---|---|---|
| | | *Baseline* | *Proposed* | *Baseline* | *Proposed* |
| Total | Extraversion | 0.685 | **0.655** | 0.730 | **0.692** |
| | Agreeableness | 0.444 | **0.399** | 0.298 | **0.262** |
| | Conscientiousness | 0.397 | **0.249** | 0.249 | **0.240** |
| | Neuroticism | 0.618 | **0.591** | 0.562 | **0.545** |
| | Openness | 0.622 | **0.619** | 0.517 | **0.515** |
| Female | Extraversion | 0.621 | **0.573** | 0.729 | **0.601** |
| | Agreeableness | 0.393 | **0.381** | 0.258 | **0.242** |
| | Conscientiousness | 0.561 | **0.492** | 0.415 | **0.334** |
| | Neuroticism | 0.612 | **0.532** | 0.554 | **0.425** |
| | Openness | 0.709 | 0.709 | 0.625 | 0.625 |
| Male | Extraversion | 0.691 | **0.661** | 0.736 | **0.734** |
| | Agreeableness | 0.422 | **0.415** | 0.270 | **0.264** |
| | Conscientiousness | 0.407 | **0.393** | 0.293 | **0.275** |
| | Neuroticism | 0.571 | **0.525** | 0.536 | **0.463** |
| | Openness | 0.521 | **0.520** | 0.400 | 0.400 |

On the other hand, we found that the performance of neuroticism prediction is better in the gender-specific model than in the entire-population model, which may due to the different key features for male and female. According to our explanations in Section 4.1, male and female with high neuroticism scores may have very different regularity of activity intensity in the night. However, if we do not consider the gender difference, the regularity of activity intensity will not become the key features in the entire population. This phenomenon addresses the importance of building the gender-specific prediction models for the neuroticism personality trait.

Our model is less effective in predicting openness trait. The reason may be of that human physical activity intensity is not strongly associated with the openness trait. In daily life, it is also hard to tell if someone is inventive or curious by his/her activity intensity pattern.

For the current research, there are some limitations which need to be addressed in the future. Firstly, the sample size of our adopted dataset (52) is relatively small, which may limit the performance of personality prediction. Further research is needed to explore larger dataset to prove the effectiveness of physical activity features. Secondly, the evaluation method is relatively simple and a comprehensive evaluation method needs to be proposed for better comparing with existing work. Lastly, the existence of biases in the Big-5 self-report data such as sampling bias, response biases (i.e., misunderstanding of the measurement, social desirability bias, 'look good' in the survey) may affect prediction performance. Further work needs to recognize and mitigate such biases.

## CONCLUSION

In this research, we first demonstrated that it is possible to combine human physical activity intensity data with traditional phone activity data to estimate the Big-5 personality traits score. We proposed a set of important metrics based on dispersion, diversity, regularity, etc., and found some interesting associations between human activity patterns and personality traits. Support Vector Regression is used to predict participants' personality scores.

Experiment results show that our predictive model is highly correlated with the ground truth and outperforms the baseline model. We also found that the performance of our predictive model differs in the female and male group, with observable reduction of errors compared with the total participants' group.

This research presents a significant step to passive human personality prediction from the measurements of smartphone activity data. In the future, larger datasets will be explored to prove the effectiveness of physical activity features in personality prediction. Different activity types will be extracted to enhance our predictive model. Also, recognizing and mitigating the biases in the Big-5 self-report data will become another important direction in our future research.

## ACKNOWLEDGMENTS

This research was supported by the Australian Government through the Australian Research Council's Linkage Projects funding scheme (project LP150100246).

# AUTHOR BIOS

Nan Gao is a PhD candidate in School of Science, RMIT University, Melbourne, Australia. Her research interests include human behaviour analytics and personality prediction. Email: nan.gao@rmit.edu.au

Wei Shao received the Ph.D. degree at RMIT University, Melbourne, Australia in 2018. His current research interests include spatio-temporal data analysis, and device-free activity recognition. Email: wei.shao@rmit.edu.au

Flora Salim received her Ph.D. in 2009 from Monash University. She is an Associate Professor and the Deputy Director of Centre for Information Discovery and Data Analytics, RMIT University. Her research interests are human mobility and behaviour analytics, context and activity recognition, and urban intelligence. She received Alexander von Humboldt Fellowship in 2019, and Victoria Fellowship in 2018. She is an Associate Editor of the PACM on Interactive, Mobile, Wearable and Ubiquitous Technologies (IMWUT). She is an Area Editor of Pervasive and Mobile Computing. Email: flora.salim@rmit.edu.au